\documentclass[twocolumn,pre,superscriptaddress,10pt,aps,floatfix]{revtex4-1}

\usepackage[nice]{nicefrac}
\usepackage{graphicx,color}
\usepackage{amsmath,amssymb,bm,bbm}
\usepackage{amsmath}
\usepackage{braket}
\usepackage{bbold}
\usepackage{float}
\usepackage{physics}
\usepackage{xcolor}
\usepackage{duckuments}
\usepackage[T1]{fontenc}

\usepackage[plainpages=false,pdfpagelabels,colorlinks=true,linkcolor=red,urlcolor=blue,citecolor=blue,pdftitle={},pdfauthor={},pdfdisplaydoctitle=true,pdfduplex=DuplexFlipLongEdge]{hyperref}

\definecolor{darkred}{rgb}{0.90,0.2,0.2}
\definecolor{darkgreen}{rgb}{0,0.60,.2}
\definecolor{darkblue}{rgb}{0.1,0.3,1}
\definecolor{grey}{cmyk}{0,0,0,0.25}
\definecolor{orange}{cmyk}{0,0.6,0.8,0}

\definecolor{max}{HTML}{03A678}
\definecolor{rafal}{HTML}{161226}
\definecolor{lev}{HTML}{0312A8}
\definecolor{miroslav}{HTML}{1312A8}

\begin{document}

\title{Fading ergodicity}

\author{Maksymilian Kliczkowski}
\affiliation{Department of Physics, Faculty of Mathematics and Physics, University of Ljubljana, SI-1000 Ljubljana, Slovenia\looseness=-1}
\affiliation{Department of Theoretical Physics, J. Stefan Institute, SI-1000 Ljubljana, Slovenia}
\affiliation{Institute of Theoretical Physics, Faculty of Fundamental Problems of Technology, Wrocław University of Science and Technology, 50-370 Wrocław, Poland\looseness=-3}
\affiliation{These authors contributed equally: Maksymilian Kliczkowski, Rafał Świętek\looseness=-4}
\author{Rafał Świętek}
\affiliation{Department of Theoretical Physics, J. Stefan Institute, SI-1000 Ljubljana, Slovenia}
\affiliation{Department of Physics, Faculty of Mathematics and Physics, University of Ljubljana, SI-1000 Ljubljana, Slovenia\looseness=-1}
\affiliation{These authors contributed equally: Maksymilian Kliczkowski, Rafał Świętek\looseness=-4}
\author{Miroslav Hopjan}
\affiliation{Department of Theoretical Physics, J. Stefan Institute, SI-1000 Ljubljana, Slovenia}
\author{Lev Vidmar}
\affiliation{Department of Theoretical Physics, J. Stefan Institute, SI-1000 Ljubljana, Slovenia}
\affiliation{Department of Physics, Faculty of Mathematics and Physics, University of Ljubljana, SI-1000 Ljubljana, Slovenia\looseness=-1}

\begin{abstract}
Eigenstate thermalization hypothesis (ETH) represents a breakthrough in many-body physics since it allows to link thermalization of physical observables with the applicability of random matrix theory (RMT).
Recent years were also extremely fruitful in exploring possible counterexamples to thermalization, ranging, among others, from integrability, single-particle chaos, many-body localization, many-body scars, to Hilbert-space fragmentation.
In all these cases the conventional ETH is violated.
However, it remains elusive how the conventional ETH breaks down when one approaches the boundaries of ergodicity, and whether the range of validity of the conventional ETH coincides with the validity of RMT-like spectral statistics.
Here we bridge this gap and we introduce a scenario of the ETH breakdown in many-body quantum systems, dubbed fading ergodicity regime, which establishes a link between the conventional ETH and non-ergodic behavior.
We conjecture this scenario to be relevant for the description of finite many-body systems at the boundaries of ergodicity, and we provide numerical and analytical arguments for its validity in the quantum sun model of ergodicity breaking phase transition.
For the latter, we provide evidence that the breakdown of the conventional ETH is not associated with the breakdown of the RMT-like spectral statistics.
\end{abstract}

\maketitle

\section{Introduction}

The quantum chaos conjecture links the emergence of random-matrix theory (RMT) statistics in quantum systems with the chaotic dynamics in their classical limit~\cite{bohigas_giannoni_84, casati_valzgris_80}.
Perhaps surprisingly or not, the RMT predictions are also relevant for the description of spectral statistics of quantum many-body systems without classical counterparts, e.g., of interacting spin-1/2 systems on a lattice~\cite{montambaux_poilblanc_93, hsu_dauriac_93, distasio_zotos_95, prosen_99, santos_04, rabson_narozhny_04, rigol_santos_10}.
Nevertheless, the experiments that study nonequilibrium dynamics of isolated quantum many-body systems usually cannot access the spectral properties, but they can measure local observables such as site occupations~\cite{bloch_dalibard_12}.
Here, the central role is played by the eigenstate thermalization hypothesis (ETH)~\cite{deutsch_91, srednicki_94, rigol_dunjko_08}, which provides simple principles to explain the agreement between the observable expectation values in time-evolving pure states and the predictions of statistical ensembles~\cite{dalessio_kafri_16}.

The possibility for thermalization to occur on a level of eigenstates is suggested by the analysis of expectation values of observables in random pure states~\cite{dalessio_kafri_16}.
However, Hamiltonian eigenstates are not random pure states and hence the ETH contains non-trivial refinements beyond the RMT.
In particular, considering the expectation values of an observable $\hat O$ in Hamiltonian eigenstates, $\hat H|n\rangle = E_n|n\rangle$, the non-trivial refinements represent the structure function $O(\bar E)$ of the diagonal matrix elements, where $\bar E = (E_n+E_m)/2$ is the mean energy, and the envelope function $f(\bar E,\omega)$, where $\omega=E_n-E_m$ is the energy difference (setting $\hbar \equiv 1$).
Combining these with the fluctuating part that originates from the analysis of random pure states gives rise to the ETH ansatz (also referred to as the {\it conventional} ETH further on)~\cite{srednicki_99},
\begin{equation} \label{def_eth}
    \langle n| \hat O |m\rangle = O(\bar E) \delta_{m,n} + \rho(\bar E)^{-1/2} f(\bar E,\omega) R_{nm} \;.
\end{equation}
In the latter, the fluctuations are suppressed as a square root of the many-body density of states $\rho(\bar E)$ that increases exponentially with the number of lattice sites $L$ in interacting systems, and $R_{nm}$ is a random number with zero mean and variance one.
Equation~(\ref{def_eth}) provides a mechanism of thermalization in an isolated quantum system for which short-range statistics agree with RMT predictions~\cite{dalessio_kafri_16}.

When considering counterexamples to thermalization, the ETH from Eq.~(\ref{def_eth}) is not expected to be valid.
Specifically, two features of the matrix elements may emerge in non-ergodic systems: 
(a) the fluctuations of matrix elements may decay polynomially (instead of exponentially) with $L$, and (b) the matrix elements in some eigenstates, dubbed outliers, may not approach the corresponding microcanonical averages.
These frameworks allow for introducing weaker forms of ETH that apply, e.g., for integrable systems~\cite{biroli_lauchli_10, ikeda_ueda_15, alba_15, leblond_mallayya_19, brenes_leblond_20}, single-particle chaotic systems~\cite{lydzba_swietek_24}, many-body localization~\cite{luitz_16, panda_scardicchio_19, Colmenarez2019, corps_molina_21}, many-body scars~\cite{shiraishi_mori_17, *mondaini_mallayya_18, serbyn_abanin_2021, moudgalya_2022}, and Hilbert-space fragmentation~\cite{sala_rakovszky_2020, khemani_hermele_2020, moudgalya_2022}.
However, all these forms of ETH are incompatible with ergodicity and thermalization. 
These considerations give rise to the central question of our study: how and when does the conventional ETH from Eq.~(\ref{def_eth}) evolve into other (weaker) forms of ETH when ergodicity is fading, and to what extent is this transition related to the breakdown of RMT-like short-range spectral statistics?

It is intriguing that, in contrast to observables, the approach towards boundaries of ergodicity is currently better understood for spectral properties.
In physical systems, one can define the Thouless energy $\Gamma$ that separates the properties of short-range and long-range spectral statistics.
The short-range statistics comply with the RMT predictions while the long-range do not, and $\Gamma$ shrinks to the mean level spacing (i.e., the Heisenberg energy) at the ergodicity breaking transition.
This perspective is well established for the Anderson localization transition~\cite{sierant_delande_20, suntajs_prosen_21}, and also for certain well controlled many-body systems such as the quantum sun model of the avalanche theory~\cite{suntajs_vidmar_22}. 
The inverse of the Thouless energy, dubbed Thouless time $t_{\rm Th}$, has a natural interpretation of a diverging relaxation time at the ergodicity breaking phase transition.

Here, our goal is to go beyond the paradigm of ergodicity breaking as the absence of validity of RMT spectral statistics, and hence we focus our analysis on observables.
We introduce a scenario in which the fluctuations of the observable matrix elements soften when approaching the ergodicity boundary, even though the observables still thermalize and the short-range spectral statistics comply with RMT predictions.
Our theory provides a natural bridge between the two limits, i.e., the conventional ETH limit~(\ref{def_eth}) and the completely non-ergodic limit.
Moreover, in the regime intermediate to these two limits, dubbed fading ergodicity regime, it establishes ergodicity beyond the conventional ETH.
Hence, it can be used as observable-based precursor of the ergodicity breaking phase transition.

\section{Scenarios for the ETH breakdown}

We now discuss the possible scenarios for the ETH breakdown in many-body quantum systems.
The latter should be searched for observables that are diagonal in the basis in which the Hamiltonian eigenstates in the non-ergodic regime exhibit signatures of localization.
This perspective is supported by the recent analysis of the matrix elements of observables in single-particle eigenstates of an Anderson insulator~\cite{lydzba_zhang_21}.
As the limiting non-ergodic behavior, we hence expect the majority of the weight of the matrix elements to be accumulated in the diagonal matrix elements.
An important technical aspect when considering ETH is the existence of a sum rule for the matrix elements of observables~\cite{lydzba_swietek_24}, which we impose as
\begin{equation} \label{def_sumrule}
\frac{1}{\cal D}\sum_{n,m=1}^{\cal D}|O_{nm}|^2 = 1\;,
\end{equation}
where $\cal D$ is the Hilbert space dimension and $O_{nm} \equiv \langle n|\hat O|m\rangle$, see Appendix~\ref{sec:sm:matrix_elems} for details.

Based on these considerations, we anticipate two possible scenarios when approaching the ergodicity boundary:
(a) the conventional ETH, as given by Eq.~(\ref{def_eth}), is valid in the entire ergodic phase, i.e., its validity coincides with the short-range spectral statistics being RMT-like;
(b) the deviations from Eq.~(\ref{def_eth}), e.g., in the form of softening of the fluctuations, are manifested despite the system being ergodic and the short-range level statistics complying with the RMT predictions.

We note that recent studies based on the norms of adiabatic gauge potentials contributed valuable insights into the structure of the off-diagonal matrix elements in the vicinity of the ergodicity breakdown~\cite{pandey_claeys_20}.
In particular, they suggested that the softening of fluctuations of matrix elements at low $\omega$ is s smooth process in the vicinity of the breakdown~\cite{leblond_sels_21, kim_polkovnikov_24}.
However, a clear distinction between the scenarios (a) and (b) has to our knowledge not yet been established.

\begin{figure}[t!]
\centering
\includegraphics[width=\columnwidth]{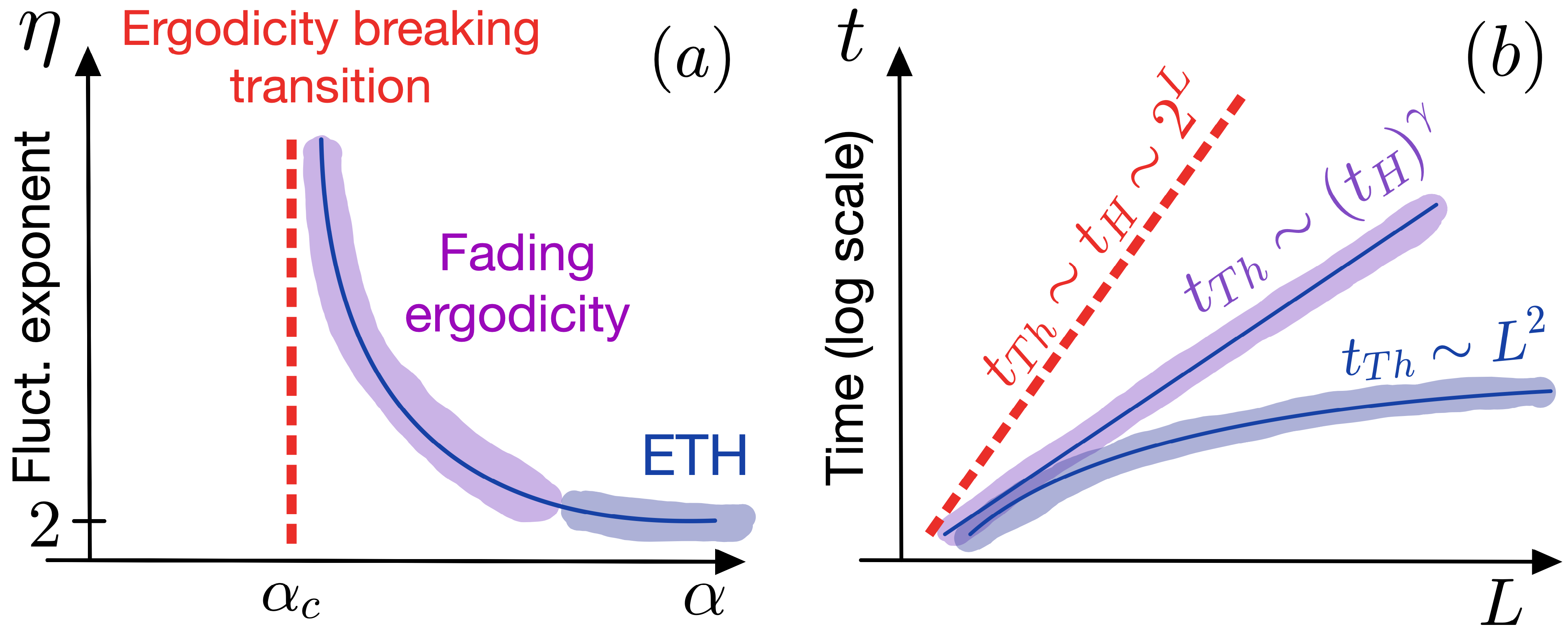}
\caption{Sketch of the fading ergodicity scenario. 
(a) Divergence of the fluctuation exponent $\eta$, see Eq.~(\ref{def_eta}), as a function of the control parameter $\alpha$, when approaching the ergodicity breaking transition point at $\alpha=\alpha_c$. 
(b) While the Thouless time $t_{\rm Th}$ is proportional to the Heisenberg time $t_{H}$ at the transition point ($t_{\rm Th} \sim t_H$), and it is much smaller than $t_H$ in the conventional ETH regime (e.g., $t_{\rm Th} \sim L^2$), it scales as $t_{\rm Th} \sim (t_H)^\gamma$, with $0<\gamma<1$, when the boundary of ergodicity is approached.
}
\label{fig1}
\end{figure}

Here we put forth arguments in favor of scenario (b).
Our main conjecture is that the fluctuations of the diagonal and low-$\omega$ off-diagonal matrix elements soften when approaching the ergodicity boundary, such that the fluctuating part of the ETH ansatz in Eq.~(\ref{def_eth}) should acquire $\omega$-dependence, $\rho(\bar E)^{-1/2} \to \Sigma(\bar E,\omega,L)$.
Hence, the difference of our scenario with respect to the conventional ETH is that the $L$-dependence of fluctuations, expressed via the density of states $\rho(\bar E)$, and the $\omega$-dependence of the off-diagonal matrix elements, do not decouple as in Eq.~(\ref{def_eth}).
In the low-$\omega$ regime $\omega\to 0$, we express the softening of fluctuations as
\begin{equation} \label{def_eta}
    \rho(\bar E)^{-1/2} \;\; \to \;\; \Sigma(\bar E,\omega\to 0, L) \;\; \to \;\;  \rho(\bar E)^{-1/\eta}\;,
\end{equation}
with $2 < \eta < \infty$ in the fading ergodicity regime.
We propose that $\eta$ is determined by the ratio of the Thouless and Heisenberg energy, and when the Thouless time is sufficiently large, the divergence of $\eta$ signals proximity of the ergodicity breaking phase transition. 
This scenario is sketched in Fig.~\ref{fig1}.

\section{Softening of ETH at small energies}

We now present an argument supporting Eq.~(\ref{def_eta}), i.e., a scenario that naturally interpolates between the conventional ETH ($\eta=2$) and its breakdown in the non-ergodic limit ($\eta\to\infty$).
The argument is based on the $L$-dependence of the off-diagonal matrix elements at low $\omega$ for the targeted observable $\hat O$.
For simplicity, we model the coarse-grained off-diagonal matrix elements at low $\omega$ with a Lorentzian function.
This specific functional form is not crucial for the derivation of our main result, however, since we are approaching the ergodicity boundary from the ergodic side, it is important that the function is a constant in the $\omega\to 0$ limit.
We consider
\begin{equation} \label{def_Oab_scaled_1}
    \overline{|O_{nm}|^2} \rho = \frac{\Gamma}{\Gamma^2 + \omega^2}  \;,
\end{equation}
where for simplicity we omit the energy dependence of $\rho$.
The width of the function $\Gamma$ is a characteristic low-energy scale that we name the Thouless energy.
We refer to low-$\omega$ off-diagonal matrix elements as those that belong to the interval $\Delta < \omega < \Gamma$, where $\Delta \propto 1/\rho$ is the many-body mean level spacing, i.e., the Heisenberg energy.

Importantly, the Thouless energy $\Gamma$ depends on $L$ in most physical systems (it vanishes as $L\to\infty$). 
Consequently, also the scaled off-diagonal matrix elements $\overline{|O_{nm}|^2} \rho$ are expected to depend on $L$.
This is a known property, which is usually interpreted via the $L$-dependence of $f(\bar E,\omega)$, see Eq.~(\ref{def_eth}), at $\omega \ll \Gamma$~\cite{dalessio_kafri_16, luitz_barlev_16, brenes_leblond_20, richter_dymarsky_20, schoenle_jansen_21, wang_lamann_22, Sugimoto2022, balachandran_santos_23}.
However, in many cases this $L$-dependence is polynomial, and it is hence subleading when compared to the exponentially increasing $\rho$.

Equation~(\ref{def_Oab_scaled_1}) suggests that one may restore scale invariance by expressing the matrix elements as a function of $\omega/\Gamma$, giving rise to the following scale-invariant form:
\begin{equation} \label{def_Oab_restored}
\overline{|O_{nm}|^2} \rho \Gamma = \frac{1}{1 + (\omega/\Gamma)^2}.    
\end{equation}
As a consequence, the low-energy offdiagonals at $\omega \ll \Gamma$ scale as
\begin{equation} \label{def_Oab_modified}
    \overline{|O_{nm}|^2} \propto \frac{1}{\rho \Gamma} \approx \frac{\Delta}{\Gamma} \;.
\end{equation}
Equation~(\ref{def_Oab_modified}) is the main result of this study: whenever the Thouless energy increases as $\Gamma \propto \Delta^\zeta$, with $0<\zeta<1$, the system is still ergodic but the ETH does not hold in the conventional way~(\ref{def_eth}).
In particular, the fluctuations of the matrix elements are softened according to Eq.~(\ref{def_eta}) as $\eta = 2/(1-\zeta)>2$, i.e., $\eta$ diverges at the ergodicity breaking transition at which the Thouless energy scales as the Heisenberg energy, $\Gamma \propto \Delta$.

The physical picture that emerges from our study is the suppression of fluctuations of the low-$\omega$ off-diagonal matrix elements, as well as the diagonal matrix elements, which can be interpreted as the accumulation of spectral weight in the ergodic system at $\omega\ll \Gamma$.
In contrast, the sum rule for the matrix elements from Eq.~(\ref{def_sumrule}) then suggests the depletion of the spectral weight at $\omega \gg \Gamma$.
We expect this phenomenology to be a generic feature of finite ergodic systems that approach the boundaries of ergodicity, irrespective of whether the ergodicity breaking transition becomes a true phase transition in the thermodynamic limit, or the non-ergodic regime ultimately becomes a singular point in the parameter space.

\section{Example: Quantum sun model}

We now present analytical and numerical evidence that supports the above scenario in a toy model that hosts an ergodicity breaking phase transition in the thermodynamic limit, i.e., the quantum sun model of the avalanche theory~\cite{deroeck_huveneers_17, luitz_huveneers_17, suntajs_vidmar_22}. The Hamiltonian is defined as
\begin{equation} \label{def_qsm}
    \hat H = \hat{H}_{\rm dot} 
    + \sum _ {j=1} ^ {L} \alpha ^{u_j} \hat {S} _{n(j)}^x \hat {S}_j ^{x} 
    + \sum _{j=1}^{L} h_j \hat{S}_j^z,
\end{equation}
where $\hat{H}_{\rm dot}$ is a $2^N \times 2^N$ random matrix drawn from the Gaussian orthogonal ensemble (GOE) describing all-to-all interacting particles within an ergodic quantum dot (we set $N=3$).
The second term defines the coupling between a spin $j$ outside the dot ($j=1,...,L)$ and a randomly selected spin $n(j)$ within the dot, with $\alpha$ acting as the tuning parameter of the ergodicity breaking phase transition, and $u_j \propto j$.
Details of the model implementation and the choice of model parameters are given in Appendix~\ref{sec:sm:model}.

A convenient aspect of the quantum sun model is that it allows for using accurate closed-form expressions for the Thouless and Heisenberg energies.
The Thouless energy on the ergodic side can be well approximated as~\cite{suntajs_vidmar_22}
\begin{equation}
    \Gamma \propto e^{-\ln\left(\frac{1}{\alpha^2}\right)L}\;,
\end{equation}
while the Heisenberg energy scales, up to subleading corrections, as
$\Delta \propto 2^{-L} = \exp{-L\ln{\qty(1/\tilde{\alpha}_c^2)}}$, where $\tilde{\alpha}_c = 1/\sqrt{2}$ is the critical point derived within the avalanche theory~\cite{deroeck_huveneers_17, suntajs_hopjan_24}.
Hence, using Eq.~(\ref{def_Oab_modified}), one can estimate the scaling of the low-$\omega$ off-diagonal matrix elements as
\begin{equation}
    \overline{|O_{nm}|^2} \propto e^{-\ln\left(\frac{\alpha^2}{\tilde{\alpha}_c^2}\right)L}\;,
\end{equation}
which leads to a closed-form expression for the divergence of $\eta$ from Eq.~(\ref{def_eta}),
\begin{equation}\label{eq:eta_analytic}
    \eta^* = 2 \left( 1 - \frac{\ln\alpha}{\ln\tilde{\alpha}_c} \right)^{-1}\;,
\end{equation}
in the ergodic phase at $\tilde{\alpha}_c \leq \alpha \leq 1$.
While $\tilde{\alpha}_c = 1/\sqrt{2} \approx 0.707$ provides a reasonably accurate prediction of the transition point, in what follows we refer to the numerically extracted transition point as $\alpha_c$, and the difference between $\tilde{\alpha}_c$ and $\alpha_c$ is of the order of a few percents.

\begin{figure}[!t]
\centering
\includegraphics[width=\columnwidth]{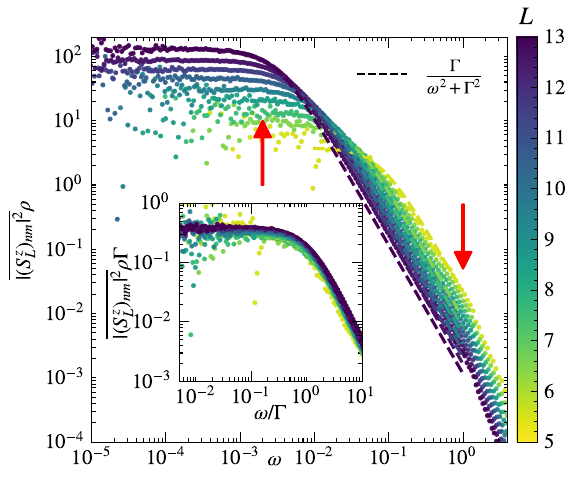}
\caption{
Coarse-grained off-diagonal matrix elements of $\overline{|(S_L^z)_{nm}|^2}$ at $\alpha = 0.86$ and different $L$.
Main panel: $\overline{|(S_L^z)_{nm}|^2} \rho$ vs $\omega$, where the density of states $\rho\propto 2^L$ is obtained from a small energy window in the middle of the spectrum.
Dashed lines are fits to the Lorentzian function~(\ref{def_Oab_scaled_1}), from which we extract $\Gamma$.
The red arrows highlight the weight accumulation (depletion) at low (high) $\omega$. 
Inset: $\overline{|(S_L^z)_{nm}|^2} \rho\Gamma$ vs $\omega/\Gamma$.
}
\label{fig2}
\end{figure}

Below we numerically test the above predictions. We focus on the observable $\hat O = \hat S_L^z$, i.e., the $z$-component of the most weakly coupled spin to the ergodic quantum dot.
The physical motivation for selecting this observable is based on understanding that observables, which measure properties of spins that are most weakly coupled to the ergodic quantum dot, are most sensitive to ergodicity breaking~\cite{suntajs_hopjan_24}, and hence their relaxation time can be associated with the Thouless time.
Results for other observables are shown in Appendix~\ref{sec:sm:choice}.

\subsection{Fluctuations of matrix elements}

In Fig.~\ref{fig2}, we show the $\omega$-dependence of the coarse-grained off-diagonal matrix elements $\overline{|(S_L^z)_{nm}|^2}$, where $(S_L^z)_{nm} \equiv \langle n|\hat S_L^z|m\rangle$, and the coarse graining (followed by the averaging over Hamiltonian realizations) is carried out over a narrow window above the target $\omega$.
The scaled matrix elements $\overline{|(S_L^z)_{nm}|^2} \rho$ vs $\omega$ exhibit accumulation of spectral weight at low $\omega$ and its depletion at large $\omega$, see the arrows in the main panel of Fig.~\ref{fig2}.
Remarkably, the results in Fig.~\ref{fig2} are shown for $\alpha = 0.86$, for which the nearest level-spacing statistics complies with the GOE predictions, see Fig.~\ref{fig:sm:fig1}(a) and Appendix~\ref{sec:sm:model}.
This suggests that the softening of the fluctuations of the matrix elements emerges in the fading ergodicity regime, in which the short-range statistics are still GOE-like.
Another perspective on the softening of the low-$\omega$ matrix elements is shown in the inset of Fig.~\ref{fig2}, in which we show the scaled matrix elements $\overline{|(S_L^z)_{nm}|^2} \rho \Gamma$ vs $\omega/\Gamma$, exhibiting a reasonably good collapse of the results for all system sizes under investigation.

\begin{figure}[t!]
\centering
\includegraphics[width=\columnwidth]{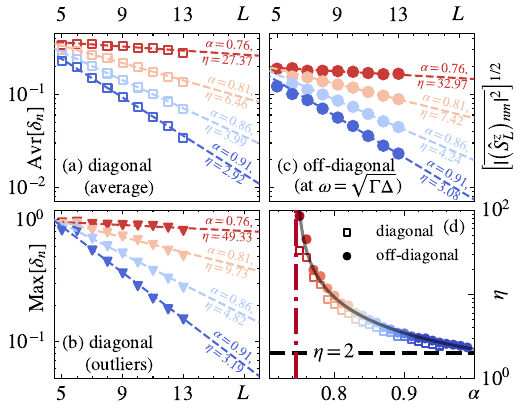}
\caption{
Scaling of fluctuations of matrix elements.
(a), (b) Eigenstate-to-eigenstate fluctuations of the diagonal matrix elements, $\delta _n \equiv |(\hat{S}^z_L)_{n+1} - (\hat{S}^z_L)_{n}|$.
We show the average and the maximal outliers in (a) and (b), respectively, see also Appendix~\ref{sec:sm:matrix_elems} for the details on the intervals of averaging.
(c) Low-$\omega$ off-diagonal matrix elements, taken from a narrow energy window around the target $\omega = \sqrt{\Gamma \Delta}$.
Dashed lines in (a)-(c) are the two-parameter fits of the function $a_0 2^{-L/\eta}$ with the values of $\eta$ shown next to them.
(d) Fluctuation exponents $\eta$ from Eq.~(\ref{def_eta}) as a function of $\alpha$.
The solid line is a fit of $b_0\eta^*$ to the results for the off-diagonal matrix elements, where $b_0=1.12$ is the fitting parameter and $\eta^*$ is the function from Eq.~(\ref{eq:eta_analytic}), in which $\tilde{\alpha}_c$ is replaced by a fitting parameter $\alpha_c$.
We get $\alpha_c = 0.745$, see the vertical dashed-dotted line, which is very close to the values of transition point obtained from other indicators such as the average gap ratio shown in Fig.~\ref{fig:sm:fig1}(b) of Appendix~\ref{sec:sm:model}.
} 
\label{fig3}
\end{figure}

A quantitative analysis of the fluctuations of the matrix elements in carried out in Fig.~\ref{fig3}.
We study the average eigenstate-to-eigenstate fluctuations of the diagonal matrix elements, see Fig.~\ref{fig3}(a), and the fluctuations of the low-$\omega$ off-diagonal matrix elements measured at $\omega = \sqrt{\Delta \Gamma}$, see Fig.~\ref{fig3}(c).
They both exhibit a fast decay deep in the ergodic regime (at $\alpha$ close to $\alpha=1$), while the decay becomes much slower when the ergodicity breaking transition point is approached, $\alpha \to \alpha_c$.
The extracted fluctuation exponents $\eta$ as functions of $\alpha$ are shown as symbols in Fig.~\ref{fig3}(d).
They behave very similarly for the diagonal and the low-$\omega$ off-diagonal matrix elements, and remarkably, are well described by the solid line that corresponds to the analytical prediction from Eq.~(\ref{eq:eta_analytic}).
Moreover, Fig.~\ref{fig3}(b) shows that even the maximal outliers of the diagonal matrix elements decay to zero with increasing $L$.
This property suggest that the system is still ergodic, and makes a clear distinction to non-ergodic systems such as integrable interacting systems in which the maximal outliers do not decay with system size~\cite{biroli_lauchli_10, leblond_mallayya_19, lydzba_mierzejewski_23}.

While the focus in the main text is on the fluctuations of matrix elements,  we also study their distributions in Appendix~\ref{sec:sm:distributions}.
There, we show evidence that the distributions appear to be close to a normal distribution in nearly entire ergodic phase.

\subsection{Quantum dynamics} \label{sec:quench}

We now demonstrate the emergence of ergodicity, and its breakdown, from the dynamical perspective.
Our main goal is to show that the fading ergodicity regime can still be characterized as ergodic despite the breakdown of the conventional ETH.
On the other hand, we show evidence that the breakdown of ergodicity occurs when the fluctuation exponent diverges, $\eta\to\infty$.

We study the dynamics of the expectation value of $\hat S_L^z$ after a sudden quantum quench.
For a single Hamiltonian realization $\hat H^{(\mu)}$, it is defined as
\begin{equation} \label{def_Qt}
    \mathcal{Q}(t)^{(\mu)} = \langle \psi_0 | \hat S_L^z(t) |\psi_0\rangle\;,
\end{equation}
where $\hat S_L^z(t) = \exp(i\hat H^{(\mu)}t) \hat S_L^z \exp(-i\hat H^{(\mu)}t)$.
The initial state $|\psi_0\rangle$ is a product state in the middle of the spectrum, see Appendix~\ref{sec:sm:time_evo} for details.
To study whether the system thermalizes or not, we first define the expectation values of $\hat S_L^z$ in the diagonal ensemble, $\mathcal{Q}_{\rm DE}^{(\mu)}$, and in the microcanonical ensemble, $\mathcal{Q}^{(\mu)} _{\rm ME}$, see Eqs.~(\ref{eq:sm:o_infty}) and~(\ref{eq:sm:eth_equivalence_time}) for definitions.
This allows us to compare the long-time value after a quantum quench to the microcanonical ensemble prediction via the quantity
\begin{equation} \label{def_dQ_infty}
    \Delta \mathcal{Q}_\infty = {\rm Avr}_\mu \left\{| \mathcal{Q}_{\rm DE}^{(\mu)} - \mathcal{Q}^{(\mu)} _{\rm ME} |\right\} \;,
\end{equation}
where ${\rm Avr}_\mu \{...\}$ corresponds to the average over Hamiltonian realizations.

\begin{figure}[t!]
\centering
\includegraphics[width=0.85\columnwidth]{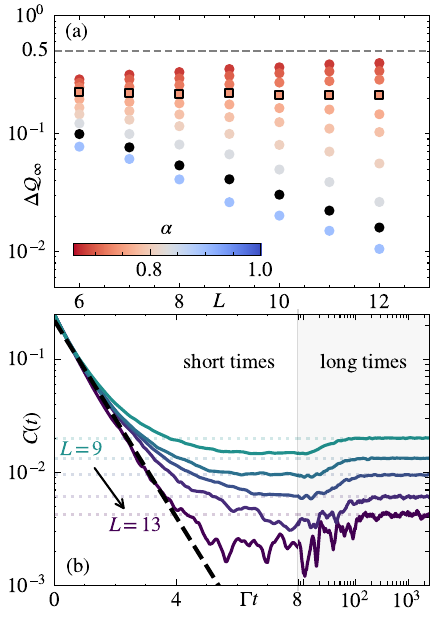}
\caption{
Quantum dynamics of the observable $\hat S_L^z$.
(a)
Difference $\Delta \mathcal{Q}_\infty$, see Eq.~(\ref{def_dQ_infty}), between the microcanonical ensemble and the diagonal ensemble prediction after a quantum quench, vs $L$.
Squares are results at $\alpha = 0.74 \approx \alpha_c$, at which the fluctuation exponent $\eta$ in Fig.~\ref{fig3}(d) diverges, $\eta\to\infty$.
Black circles are results for $\alpha=0.86$, which is studied in Fig.~\ref{fig2}.
(b)
Time evolution of the autocorrelation function $C(t)$, see Eq.~(\ref{def_Ct_autocorr}), at $\alpha = 0.86$ and different system sizes $L$, as a function of scaled time $\Gamma t$.
Dashed line denotes the exponential decay $\propto e^{-\Gamma t}$, and horizontal dotted lines represent the long-time average $C_\infty$.
}
\label{fig4}
\end{figure}

Figure~\ref{fig4}(a) shows the results for $\Delta \mathcal{Q}_\infty$ vs $L$ for different values of interaction $\alpha$.
We observe that in the fading ergodicity regime at $\alpha>\alpha_c$, the differences $\Delta \mathcal{Q}_\infty$ appear to vanish in the thermodynamic limit $L \to \infty$.
This suggests emergence of ergodicity for any finite value of the fluctuation exponent $\eta$.
On the other hand, the infinite value of $\eta$, which we observe at $\alpha<\alpha_c$, see Fig.~\ref{fig3}(d), does not ensure ergodicity.
The breakdown of ergodicity is manifested in Fig.~\ref{fig4}(a) for the results at $\alpha<\alpha_c$ (red dark circles above the squares), for which $\Delta \mathcal{Q}_\infty$ appears to approach a nonzero value in the thermodynamic limit.

Focusing on the dynamics at short and moderate times, we study a related quantity, the spin-spin autocorrelation function.
We introduce it as
\begin{equation} \label{def_Ct_autocorr}
C(t) = \langle \hat {S}_L^z (t) \hat {S}_L^z (0) \rangle_\mu \;,
\end{equation}
where the brackets $\langle\cdots\rangle_\mu$ denote both the quantum expectation value and the disorder average, see Eqs.~(\ref{def_Ct_mu}) and~(\ref{def_Ct_avr}) of Appendix~\ref{sec:sm:time_evo} for the precise definition.
The autocorrelation function $C(t)$ is shown in Fig.~\ref{fig4}(b) in the ergodic regime at $\alpha=0.86$.
It exhibits an exponential decay with the rate given by the Thouless energy $\Gamma$.
At long times it approaches the steady-state value that vanishes with increasing the system size $L$, which is consistent with ergodic behavior.

Another important property of quantum quench dynamics is whether or not the expectation values of observables, such as $\mathcal{Q}(t)^{(\mu)}$, equilibrate after a sufficiently long time.
This can be studied via the scaling of temporal fluctuations above the long-time average of expectation values.
Our main question, however, is not whether the temporal fluctuations vanish in the thermodynamic limit (they do in the entire ergodic phase), but whether their finite-size scaling exhibits softening that resembles the softening of fluctuation exponents $\eta$ shown in Fig.~\ref{fig3}(d).
The affirmative answer to this question would imply the temporal fluctuations to be an indicator, based on quantum dynamics, for the detection of fading ergodicity.
This expectation is based on their exact property (in the absence of degeneracies in the spectrum), i.e., they are upper-bounded by the decay of the maximal off-diagonal matrix element~\cite{dalessio_kafri_16}.

\begin{figure}[!t]
\centering
\includegraphics[width=1.0\columnwidth]{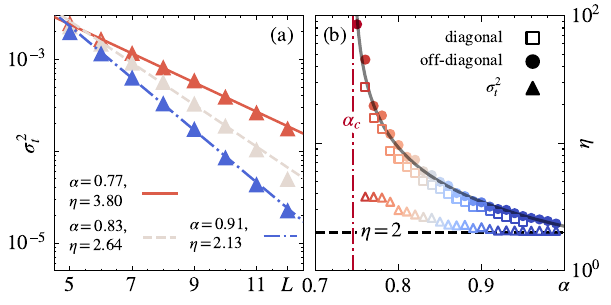}
\caption{
Scaling of the long-time temporal fluctuations.
(a)
Scaling of the variance of temporal fluctuations $\sigma_t^2$, see Eq.~(\ref{eq:sm:temportal_fluct_avr}), vs $L$, at different $\alpha$.
We extract the exponent $\eta$ of the temporal fluctuations by fitting the function $b_0 2^{-2L/\eta}$ to the numerical values (triangles).
(b) Fluctuation exponents $\eta$ vs $\alpha$.
The exponents of the temporal fluctuations are shown as triangles.
Fluctuation exponents of the diagonal (squares) and the low-$\omega$ off-diagonal matrix elements (circles) are identical to those in Fig.~\ref{fig3}(d).
}
\label{fig:sm:fig6}
\end{figure}

To address this question, we define the variance of long-time temporal fluctuations for a given Hamiltonian realization $\hat H^{(\mu)}$,
\begin{equation}\label{eq:sm:temportal_fluct}
    \sigma^2_{t,\mu} \equiv {\rm Var}_t \left\{ \mathcal{Q}(t)^{(\mu)} \right\} \;,
\end{equation}
where 
the variance ${\rm Var}_t\{\cdots\}$ is calculated in a
restricted time window after the Heisenberg time $t_{\rm H}$. Here, we select the time window $t\in[50t_H, 100t_H]$.
The corresponding average variance over the Hamiltonian realizations is defined as
\begin{equation} \label{eq:sm:temportal_fluct_avr}
    \sigma^2_{t} = {\rm Avr}_\mu \left\{ \sigma^2_{t,\mu} \right\} \;.
\end{equation}
In Fig.~\ref{fig:sm:fig6}(a) we show examples of $\sigma_t^2$ vs $L$ at different values of $\alpha$.
We observe that in the entire ergodic regime, the decay of $\sigma_t^2$ is exponential with $L$, suggesting equilibration of the system.

Results in Fig.~\ref{fig:sm:fig6}(a) motivate us to systematically study how the rate of the exponential decay of temporal fluctuations decreases when $\alpha$ approaches the ergodicity breaking transition.
In Fig.~\ref{fig:sm:fig6}(b) we compare the fluctuation exponents of temporal fluctuations with the fluctuations exponents of matrix elements fluctuations from Fig.~\ref{fig3}(d).
The obtained temporal fluctuation exponent is quantitatively smaller than the matrix element fluctuation exponent.
However, in all cases, the fluctuation exponents increase when $\alpha$ approaches the transition point $\alpha_c$, suggesting that the long-time temporal fluctuations indeed carry information about the softening of the low-$\omega$ off-diagonal matrix elements.
Hence, the breakdown of the conventional ETH in the fading ergodicity regime is also manifested in the scaling of the variance of temporal fluctuations with system size.

\begin{figure}[!t]
\centering
\includegraphics[width=0.95\columnwidth]{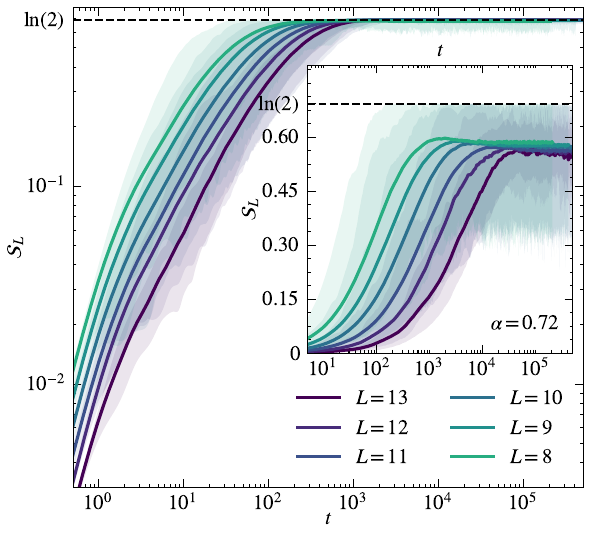}
\caption{Time evolution of the entanglement entropy ${\cal S}_L(t)$, see Eq.~(\ref{eq:sm:entropy}), of a single (most weakly coupled) spin. 
Main panel: $\alpha=0.86 > \alpha_c$ in the fading ergodicity regime.
Inset: $\alpha=0.72 < \alpha_c$ in the non-ergodic regime.
Results are shown for various system sizes $L$.
Solid lines represent the averaged entanglement entropy ${\cal S}_L(t)$, while the temporal fluctuations in different Hamiltonian realizations are denoted by fading colors.
}
\label{fig:sm:fig7}
\end{figure}

\subsection{Growth of entanglement entropy}

To complement the analysis, we study the time evolution of the entanglement entropy ${\cal S}_L(t)$.
The goal is to show that in the fading ergodicity regime, despite the softening of the fluctuations of matrix elements, one still observes the growth of the entanglement entropy towards the maximal value.
The von Neumann entanglement entropy ${\cal S}_L(t)$ is defined as
\begin{equation}\label{eq:sm:entropy}
    {\cal S}_L(t) = {\rm Avr}_\mu \left\{-\Tr{\hat{\rho}_L(t) \ln \hat{\rho}_L(t)} \right\}\;,
\end{equation}
where for each Hamiltonian realization $\hat H^{(\mu)}$, we choose the same initial state as used in Sec.~\ref{sec:quench}, and we then average the results over different Hamiltonian realizations.
In Eq.~(\ref{eq:sm:entropy}), $\hat{\rho}_L = \Tr_{\rm remain}\{|\psi(t)\rangle\langle \psi (t)|\}$ is the reduced density matrix of the single spin (at $j=L$), obtained by tracing out the remaining spins in the time-evolved state $|\psi(t)\rangle$ at time $t$.

Results shown in Fig.~\ref{fig:sm:fig7} represent the entanglement growth in the ergodic regime (main panel) and in the non-ergodic regime (inset).
We observe that in the ergodic regime, the entanglement entropy grows towards its maximum value, ${\cal S}(t\to\infty) \to \ln 2$.
The approach towards maximal entanglement entropy is consistent with the ergodic behavior of the system (however, it is not a sufficient condition; see, e.g., Ref.~\cite{vidmar_hackl_17} for a counter-example).
On the other hand, the breakdown of ergodicity is clearly manifested in the inset of Fig.~\ref{fig:sm:fig7} at $\alpha < \alpha_c$, in which the entanglement entropy after a long time saturates to a value that is below the maximal value.

\section{Conclusions}

Historically, the ETH ansatz for the matrix elements of observables was introduced by defining the smooth functions $O(\bar E)$ and $f(\bar E,\omega)$ as a refinement beyond the RMT behavior~\cite{srednicki_99, dalessio_kafri_16}.
The physical significance of this refinement was to properly describe the energy dependence of matrix elements in physical systems that exhibit thermalization.
Recent work has also explored refinements of ETH due to correlations between the matrix elements~\cite{foini_kurchan_19}, which give rise, among others, to nontrivial dynamics of four-point correlation functions~\cite{foini_kurchan_19, chan_deluca_19, murthy_srednicki_19b, pappalardi_foini_22, Dymarsky2022}.
In this work we introduce a new modification of the RMT framework that concerns the softening of the fluctuations of matrix elements.
We refer to this phenomenon as fading ergodicity, which is characterized by the breakdown of the conventional ETH.
Its physical significance is to allow for detecting boundaries of ergodicity.
In case of a well-defined ergodicity breaking phase transition in the thermodynamic limit, fading ergodicity can be considered as a precursor of the transition point at which the ETH ceases to be valid.

We provided analytical arguments and showed numerical evidence in the quantum sun model of ergodicity breaking transitions that the breakdown of conventional ETH is not associated to the breakdown of GOE-like spectral statistics, since the former occurs when later is still valid.
Yet, we argued that the unconventional form of fluctuations of matrix elements (i.e., the softening of fluctuations) is still consistent with thermalization of observables.
We expect this feature to be generic and to have applications in different types of ergodicity breaking phenomena.

\acknowledgements 
We acknowledge discussions with J. Kurchan, P. \L yd\.{z}ba, M. Mierzejewski, A. Polkovnikov, M. Rigol and P. Sierant.
We acknowledge support from the Slovenian Research and Innovation Agency (ARIS), Research core funding Grants No.~P1-0044, N1-0273, J1-50005 and N1-0369.
We gratefully acknowledge the High Performance Computing Research Infrastructure Eastern Region (HCP RIVR) consortium~\cite{vega1}
and European High Performance Computing Joint Undertaking (EuroHPC JU)~\cite{vega2}
for funding this research by providing computing resources of the HPC system Vega at the Institute of Information sciences~\cite{vega3}.
Numerical calculations have also been carried out using high-performance computing resources provided by the Wrocław Centre for Networking and Supercomputing. 

\appendix

\section{Matrix elements of observables}\label{sec:sm:matrix_elems}

\begin{figure}[b]
\centering
\includegraphics[width=\columnwidth]{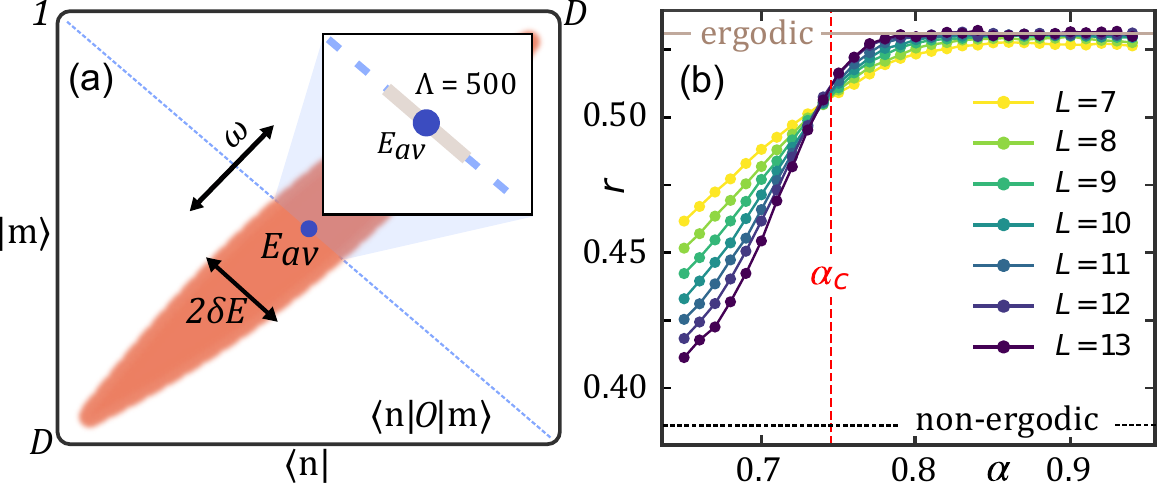}
\caption{
(a) A sketch of matrix elements $\langle n |O | m \rangle $ studied in this work.
Main panel: The shaded region represents the off-diagonal matrix elements used to extract their $\omega$-dependence.
Inset: diagonal matrix elements in the middle of the spectrum, around the mean energy $E_{\rm av}$. 
(b) Average gap ratio $r$ vs $\alpha$, for different system sizes $L$.
Vertical dashed line denotes the transition point $\alpha_c$, extracted in Fig.~\ref{fig3}(d) of the main text, while the horizontal solid and dotted lines show the ergodic and non-ergodic values, respectively.
}
\label{fig:sm:fig1}
\end{figure}

An operator matrix, expressed in the basis of the Hamiltonian eigenstates, takes the form
\begin{equation}
    \hat{O} = \sum _{n,m=1}^{\cal D} O_{nm} |n\rangle \langle m|\;,
\end{equation}
where $O_{nm}\equiv \langle n|\hat O|m\rangle$, $O_n \equiv O_{nn}$, and the states $|n\rangle$ and $|m\rangle$ denote the Hamiltonian eigenstates with corresponding eigenenergies $E_n$, satisfying $\hat{H}|n\rangle = E_{n} |n\rangle$. 
When studying ETH, it is important to properly normalize the observables~\cite{mierzejewski_vidmar_20, leblond_mallayya_19, lydzba_swietek_24}.
This is achieved, for traceless observables considered here, via the sum rule from Eq.~(\ref{def_sumrule}) in the main text.
The latter can be interpreted as requiring the Hilbert-Schmidt norm of the operator (divided by the Hilbert space dimension ${\cal D}$) to equal one.

\begin{figure*}[t!]
\centering
\includegraphics[width=0.9\textwidth]{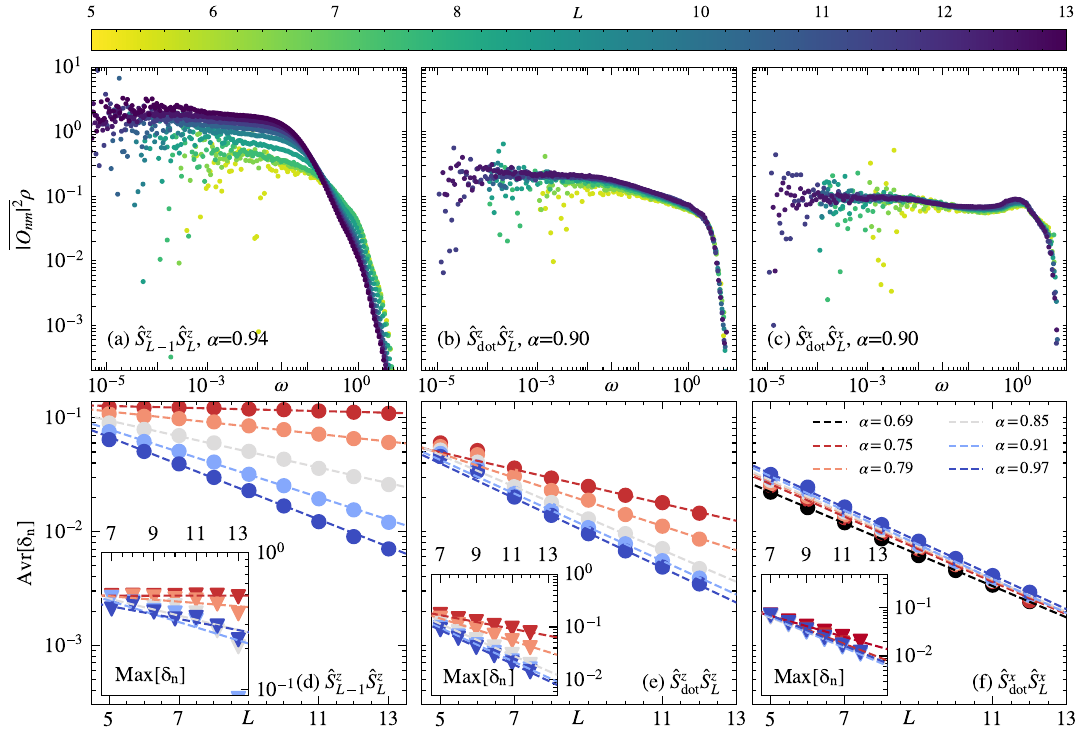}
\caption{
(a)-(c)
Coarse-grained off-diagonal matrix elements $\overline{|O_{nm}|^2}\rho$ vs $\omega$, where the density of states $\rho$ is obtained from a small energy window in the middle of the spectrum.
Results are shown for the observables $\hat{S}^z_{L-1}\hat{S}^z_{L}$, $\hat{S}^z_{\rm dot}\hat{S}^z_{L}$, and $\hat{S}^x_{\rm dot}\hat{S}^x_{L}$, respectively, for different $L$.
(d)-(f)
Eigenstate-to-eigenstate fluctuations of the diagonal matrix elements, $\delta _n \equiv |O_{n+1} - O_{n}|$, vs $L$, for the same observables as shown in (a)-(c), respectively. 
We show the average values of $\delta_n$ in the main panels and the maximal outliers in the insets.
Dashed lines are two-parameter fits to the function $a_0 2^{-L/\eta}$.
}
\label{fig:sm:fig2}
\end{figure*}

To study the diagonal matrix elements, we restrict the analysis to the subset of $\Lambda = 500$ eigenstates around the mean energy $E_{\rm av} = \Tr\{\hat{H}\} / \mathcal{D}$, as sketched in the inset of Fig.~\ref{fig:sm:fig1}(a).
For the off-diagonal matrix elements we restrict the analysis to the eigenstates within a narrow energy window $|(E_n + E_m) / 2 - E_{\rm target} | < \delta \bar {E}$ around the target energy $E_{\rm target}$, as illustrated by the shaded region in Fig.~\ref{fig:sm:fig1}(a).
We set the target energy $E_{\rm target} = E_{\rm av}$ and the energy width $\delta \bar{E}/L = 10^{-3}$, such that the ratio  of the microcanonical window to the bandwidth is $O(1)$.
In Fig.~\ref{fig2} of the main text and in Figs.~\ref{fig:sm:fig2}(a)-\ref{fig:sm:fig2}(c) we study the coarse-grained off-diagonal matrix elements $\overline{|O_{nm}|^2}$ as a function of their energy difference $\omega = E_n - E_m$.
For each bin, we obtain the values of $\overline{|O_{nm}|^2}$ by averaging over a narrow interval around the target $\omega$.
We use a fixed number of bins, which are evenly spaced on a logarithmic scale ranging from the mean level spacing $\omega \sim\Delta$ (specifically, we take $\omega=\omega_{\rm min} = 0.1/{\cal D}$) to $\omega = \omega_{\rm max}$, with $\omega_{\rm max}$ being the bandwidth of the model. 
The number of bins is of the order of hundred and it increases slightly with the system size. 

\section{Quantum sun model}\label{sec:sm:model}

The quantum sun model, defined in Eq.~(\ref{def_qsm}) of the main text, represents a system of $N+L$ spin-1/2 particles~\cite{deroeck_huveneers_17, luitz_huveneers_17, suntajs_vidmar_22, suntajs_hopjan_24}.
$N$ particles form an ergodic quantum dot with all-to-all random interactions (we fix $N=3$ throughout the study), while $L$ particles reside outside the dot.
The total Hilbert space dimension is ${\cal D} = 2^{N+L}$.
The interactions of spins within the dot are described by a $2^N\times2^N$ random matrix taken from a Gaussian orthogonal ensemble (GOE)~\cite{suntajs_hopjan_24},
\begin{equation}
    \hat{H}_{\rm dot} = \frac{1}{\sqrt{2^N + 1}} R\;,
\end{equation}
where $R = (A + A^T)/\sqrt{2}$ and the matrix elements $A_{i,j} = \mathcal{N}(0,1)$ are sampled from a normal distribution with zero mean and unit variance.
Each of the spins outside the dot is coupled to a single randomly selected spin within the dot with a strength of $\alpha ^{u_j}$, where $\alpha$ is the parameter that drives the ergodicity breaking transition, and $u_j$ (with $j=1,2,...,L$) are drawn from a uniform distribution, $u_j \in [(j - 1) - 0.2, (j - 1) + 0.2]$, except for $j=1$ when $u_1=0$.
The spins outside the dot are also subject to random magnetic fields $h_j$, which are drawn from a uniform distribution, $h_j \in [0.5, 1.5]$.
The diagonal and off-diagonal matrix elements of observables, discussed in Appendix~\ref{sec:sm:matrix_elems}, are collected using $N_{\rm samples} = 2000$ for $11\geq L \geq 5$ and $N_{\rm samples} = 874, 89$ for $L=12,13$, respectively. 

The analytical prediction for the ergodicity breaking transition point, based on the hybridization condition~\cite{deroeck_huveneers_17}, is $\tilde{\alpha}_c = 1 / \sqrt{2} \approx 0.707$.
The exact numerical calculations of different ergodicity indicators predict the value of the transition point $\alpha_c$ that is close to $\tilde{\alpha}_c$.
For the model parameters considered here the transition point is estimated to be in the interval $\tilde{\alpha}_c \lesssim \alpha_c \lesssim 0.75$~\cite{suntajs_hopjan_24}.
In particular, in Fig.~\ref{fig3}(d) of the main text we obtained $\alpha_c \approx 0.74$ using the divergence of the fluctuation exponent $\eta$ as the transition indicator.
In Fig.~\ref{fig:sm:fig1}(b) we study the standard transition indicator, namely, the average nearest level spacing ratio $r$, shortly the average gap ratio.
Introducing $r_n = {\rm min}\{\delta E_n, \delta E_{n-1}\}/{\rm max}\{\delta E_n, \delta E_{n-1}\}$, where $\delta E_n = E_{n+1}-E_n$ is the level spacing between the levels $n$ and $n+1$, we obtain the average gap ratio $r$ by averaging $r_n$ over 500 eigenstates near the center of the spectrum and over different Hamiltonian realizations.
Results in Fig.~\ref{fig:sm:fig1}(b) suggest that the scale invariant point of $r$ vs $\alpha$, which denotes the ergodicity breaking transition point~\cite{suntajs_hopjan_24}, is quantitatively very close to $\alpha_c\approx 0.74$ predicted by the analysis of $\eta$ in Fig.~\ref{fig3}(d).

\section{Choice of observables} \label{sec:sm:choice}

The analysis in the main text focuses on the observable $\hat S_L^z$, where $\hat{S}^z_j=\hat{\sigma}^z_j / 2$ and $\hat\sigma_j^z$ is a Pauli operator acting on a spin labeled by index $j$.
Here we extend the analysis to two-point spin correlations $\hat{S}^z_{L-1}\hat{S}^z_L$, $\hat{S}^z_{\rm dot}\hat{S}^z_L$ and $\hat{S}^x_{\rm dot}\hat{S}^x_L$, where $\hat{S}^z_{\rm dot}$ and $\hat{S}^x_{\rm dot}$ act on a randomly selected spin within the dot.

The results in Fig.~\ref{fig:sm:fig2} show how susceptible are the fluctuations of observable matrix elements to the proximity of the ergodicity breaking transition.
The behavior of the matrix elements of $\hat{S}^z_{L-1}\hat{S}^z_L$ is similar to the behavior of $\hat S_L^z$ studied in the main text: the spectral weight of the off-diagonal matrix elements is accumulating at low $\omega$ [cf.~Fig.~\ref{fig:sm:fig2}(a)] and the fluctuations of the diagonal matrix elements gradually soften when $\alpha$ approaches the transition point [cf.~Fig.~\ref{fig:sm:fig2}(d)].
The common property of the observables $\hat S_L^z$ and $\hat{S}^z_{L-1}\hat{S}^z_L$ is that they measure properties of spins that are at large distance from the ergodic quantum dot, and hence they exhibit the weakest coupling in the system.

On the other hand, both observables $\hat{S}^z_{\rm dot}\hat{S}^z_L$ and $\hat{S}^x_{\rm dot}\hat{S}^x_L$ include the operator within the ergodic quantum dot.
This property makes the proximity of the ergodicity breaking transition less apparent, as seen in Figs.~\ref{fig:sm:fig2}(b)-\ref{fig:sm:fig2}(c) and~\ref{fig:sm:fig2}(e)-\ref{fig:sm:fig2}(f).
For the matrix elements of $\hat{S}^z_{\rm dot}\hat{S}^z_L$, the softening of fluctuations is mild but still noticeable [cf.~Figs.~\ref{fig:sm:fig2}(b) and~\ref{fig:sm:fig2}(e)], while for $\hat{S}^x_{\rm dot}\hat{S}^x_L$, there is no signature of softening [cf.~Figs.~\ref{fig:sm:fig2}(c) and~\ref{fig:sm:fig2}(f)].
The latter is a consequence of the nonergodic phase being localized in the computational basis~\cite{suntajs_hopjan_24}, which is the eigenstate basis of the $\hat S^z$ operator.
For the observable $\hat{S}^x_{\rm dot}\hat{S}^x_L$, we find that the conventional ETH appears to be valid even on the non-ergodic side of the transition.
Hence the matrix elements of the observable $\hat{S}^x_{\rm dot}\hat{S}^x_L$ are not expected to contain any special signatures of the fading ergodicity regime.

\section{Distribution of matrix elements} \label{sec:sm:distributions}

The distributions of matrix elements of observables that comply with the conventional ETH are to a good approximation described by a normal distribution (with certain deviations that are manifested in finite systems~\cite{luitz_barlev_16, wang_wang_18, haque_mcclarty_2022, kliczkowski_swietek_23}).
We here study the distributions of the matrix elements of observables in the ergodic regime where the conventional ETH is not valid.
We ask the question to what degree are the distributions in this regime described by a normal distribution.

\begin{figure}[t!]
\centering
\includegraphics[width=\columnwidth]{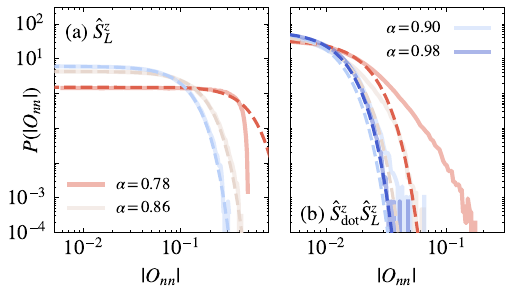}
\caption{PDFs $P(|O_{nn}|)$ of the diagonal matrix elements of observables: (a) $\hat O = \hat{S}^z_L$ and (b) $\hat O = \hat{S}^z_{\rm dot}\hat{S}^z_L$, for the system size $L=12$.
Dashed lines indicate fits to the Gaussian PDF from Eq.~(\ref{eq:sm:gaussian_pdf}). 
We numerically subtract the smooth structure function $O(\bar E)$ to study the distribution of fluctuations.
}
\label{fig:sm:fig3}
\end{figure}
\begin{figure}[t!]
\centering
\includegraphics[width=0.750\columnwidth]{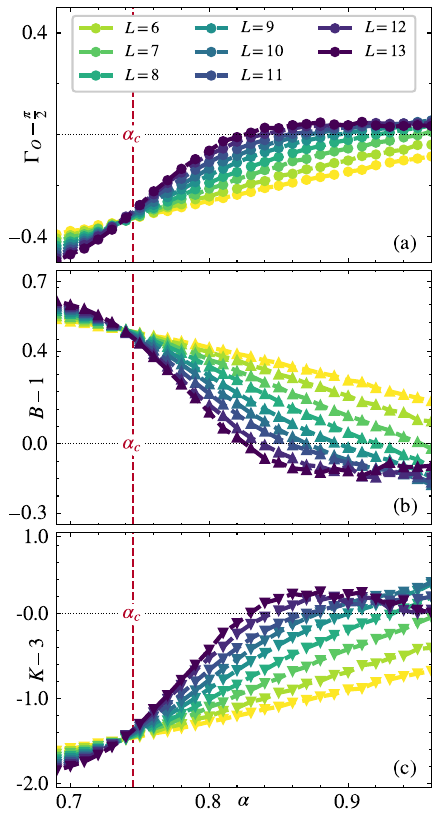}
\caption{Measures of gaussianity for the diagonal matrix elements of $\hat S_L^z$ vs $\alpha$, for different system sizes $L$.
(a) The normalized variance $\Gamma_O - \pi/2$ from Eq.~(\ref{eq:sm:gaussianity}), (b) the Binder cumulant $B-1$ from Eq.~(\ref{eq:sm:binder}), and (c) the Kurtosis $K-3$ from Eq.~(\ref{eq:sm:kurtosis}).
Horizontal dotted lines are the results for the Gaussian PDF.
Vertical dashed lines denote the transition point $\alpha_c$ obtained from Fig.~\ref{fig3}(d) in the main text.
}
\label{fig:sm:fig4}
\end{figure}
\begin{figure}[t!]
\centering
\includegraphics[width=\columnwidth]{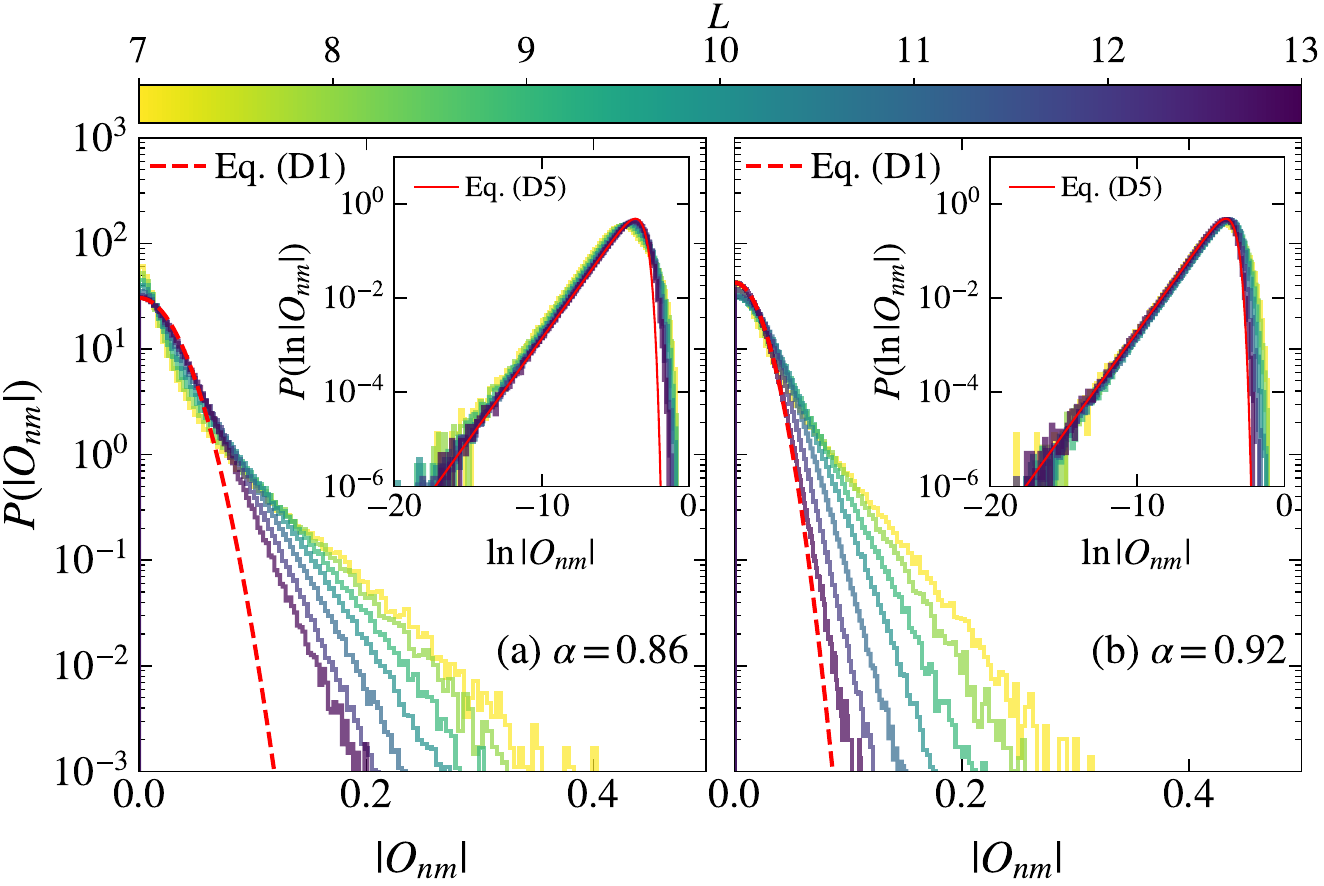}
\caption{PDFs of off-diagonal matrix elements of the observable $\hat O = \hat S_L^z$, studied in the energy interval $\omega<100 \Delta$, for (a) $\alpha=0.86$ and (b) $\alpha=0.92$.
Main panels: $P(|O_{nm}|)$ vs $|O_{nm}|$.
Dashed lines are the fits of the Gaussian PDF from Eq.~(\ref{eq:sm:gaussian_pdf}) to the largest system size.
Insets: $P(\ln|O_{nm}|)$ vs $\ln|O_{nm}|$.
Solid lines are the fits of the corresponding Gaussian PDF from Eq.~(\ref{eq:sm:gaussian_log_pdf}) using the same variance.
}
\label{fig:sm:fig3_2}
\end{figure}

Figure~\ref{fig:sm:fig3} shows the distributions of the absolute values of diagonal matrix elements of observables $\hat S_L^z$ and $\hat{S}_{\rm dot}^z\hat{S}^z_L$, for a fixed system size and for different values of $\alpha$.
Their main feature is that the distribution broadens upon decreasing $\alpha$, which is consistent with the softening of fluctuations in the regime of fading ergodicity.

The dashed lines in Fig.~\ref{fig:sm:fig3} represent the Gaussian probability density functions (PDFs).
Given a positive random variable $z$, the corresponding Gaussian PDF is given by
\begin{equation}\label{eq:sm:gaussian_pdf}
    \bar{P}(z)=\frac{2}{\sqrt{2\pi\sigma^2}}\exp{-\frac{z^2}{2\sigma^2}} \;,
\end{equation}
where a prefactor 2 appears in the numerator because we study the distribution of absolute values~\cite{kliczkowski_swietek_23}.

While the Gaussian PDFs in Fig.~\ref{fig:sm:fig3} provide accurate descriptions for the numerical distributions at large $\alpha$, they also exhibit deviations at lower $\alpha$ in the tails of the distributions.
To quantify the deviations from the Gaussian PDF as functions of both $\alpha$ and $L$, we compute the normalized variance~\cite{leblond_mallayya_19}
\begin{equation}\label{eq:sm:gaussianity}
    \Gamma_{O}=\frac{\expval{z^2}}{\expval{z}^2} \;,
\end{equation}
which yields $\Gamma_O = \pi/2$ for a Gaussian PDF from Eq.~\eqref{eq:sm:gaussian_pdf}.
We complement this measure by computing the Binder cumulant
\begin{equation}\label{eq:sm:binder}
    B=\frac{\expval{z^4}}{3\expval{z^2}^2} \;,
\end{equation}
which yields $B=1$ for the Gaussian PDF, and the Kurtosis
\begin{equation}\label{eq:sm:kurtosis}
    K=\frac{\expval{(z - \expval{z})^4}}{\expval{(z-\expval{z})^2}^2} \;,
\end{equation}
which yields $K=3$ for the Gaussian PDF. 

Figure~\ref{fig:sm:fig4} shows the dependence of the gaussianity measures from Eqs.~(\ref{eq:sm:gaussianity})-(\ref{eq:sm:kurtosis}) on $\alpha$, for the observable $\hat S_L^z$ and different system sizes $L$.
We observe that the PDFs are indeed close to a Gaussian PDF at large $\alpha$, and notable deviations occur when the transition point is approached.
However, the dependence on system size shares certain similarities with other ergodicity measures such as the average gap ratio $r$ studied in Fig.~\ref{fig:sm:fig1}(b).
This suggests a possibility that in the thermodynamic limit $L\to\infty$, the PDFs are close to Gaussian in the entire ergodic phase.
The vertical dashed line in Fig.~\ref{fig:sm:fig4} represents the critical value $\alpha_c$ obtained by fitting the divergence of the fluctuation exponent $\eta$ in Fig.~\ref{fig3} using Eq.~(\ref{eq:eta_analytic}).
Notably, all the gaussianity measures studied in Fig.~\ref{fig:sm:fig4} exhibit a scale invariant point that is very close to the extracted value of $\alpha_c$ from Fig.~\ref{fig3}.

In Fig.~\ref{fig:sm:fig3_2} we also study the distributions of the off-diagonal matrix elements at fixed $\alpha$ and different system sizes $L$.
The dashed line, which is the prediction of the Gaussian PDF from Eq.~(\ref{eq:sm:gaussian_pdf}) for the larger system size, exhibits a good agreement close the peak of the PDFs, while deviations can be observed in the tails of the distributions.
In the insets of Fig.~\ref{fig:sm:fig3_2}, we also consider the distribution of $\ln{\abs{O_{mn}}}$. 
For a random variable $x=\ln{z}$, where $z$ is distributed according to Eq.~(\ref{eq:sm:gaussian_pdf}), its distribution [shown as solid lines in the insets of Fig.~\ref{fig:sm:fig3_2}] takes the form
\begin{equation}\label{eq:sm:gaussian_log_pdf}
    \tilde{P}(x)=\frac{2e^x}{\sqrt{2\pi\sigma^2}}\exp{-\frac{e^{2x}}{2\sigma^2}} \;.
\end{equation}
The results from the insets of Fig.~\ref{fig:sm:fig3_2} reinforce the observation from the main panels that the deviations from the Gaussian PDF occur mostly in the tails, i.e., for large values of matrix elements.
The deviations are larger in Fig.~\ref{fig:sm:fig3_2}(a) [at $\alpha=0.86$] than in Fig.~\ref{fig:sm:fig3_2}(b) [at $\alpha=0.92$], suggesting similar finite-size phenomenology as observed in Fig.~\ref{fig:sm:fig4} for the diagonal matrix elements.

\section{Details on quantum dynamics}\label{sec:sm:time_evo}

In Sec.~\ref{sec:quench} of the main text, we studied the dynamics of the expectation value of observable $\hat S_L^z$ after a quantum quench, see Eq.~(\ref{def_Qt}).
We choose the initial state $|\psi_0\rangle$ such that we minimize the finite-size effects due to the existence of a many-body mobility edge in the quantum sun model~\cite{pawlik_sierant_2024}.
Specifically, for a given Hamiltonian realization, denoted as $\hat H^{(\mu)}$, we select a single initial state $|\psi_0\rangle $ that is a product state in the computational basis $|i\rangle$ (i.e., an eigenstate of $\hat S_L^z$), such that the energy of this state, $\epsilon _i= \langle i | \hat{H}^{(\mu)} | i \rangle$, is closest to the mean energy, $\epsilon _i \approx E_{\rm av}^{(\mu)}$. 
Formally, Eq.~(\ref{def_Qt}) can be expressed as
\begin{equation} \label{def_Qt_mu}
    \mathcal{Q}(t)^{(\mu)} = \langle \psi_0 | e^{iH^{(\mu)} t} \hat{S}_L^z e^{-iH^{(\mu)} t} | \psi_0\rangle\;,
\end{equation}
and further rewritten as
\begin{equation}\label{eq:sm:time_evo}
    \mathcal{Q}(t)^{(\mu)} = \sum _{n,m} c_n^{(\mu)*} c_m^{(\mu)} (S_L^z)_{nm} e^{-i(E_m^{(\mu)}-E_n^{(\mu)})t}\;,
\end{equation}
where the coefficients $\{c_n^{(\mu)}\}$ are obtained from the overlaps between the initial state $|\psi_0\rangle$ and the Hamiltonian eigenstates, $c_n^{(\mu)} = \langle n^{(\mu)}|\psi_0\rangle$. 

In the long-time limit, the expectation value approaches predictions of the diagonal ensemble (provided the spectrum has no degeneracies)~\cite{rigol_dunjko_08},
\begin{equation} \label{eq:sm:o_infty}
    \mathcal{Q}_{\rm DE}^{(\mu)} \equiv \lim _{t\rightarrow \infty} \mathcal{Q}(t)^{(\mu)} 
    =\sum _{n=1}^{\mathcal{D}} |c_n^{(\mu)}|^2 (S_L^z)_{n}\;.
\end{equation}
The expectation values of observables in the diagonal ensemble depend on the initial state of the system via the coefficients $|c_n^{(\mu)}|^2$.
We compare the diagonal ensemble prediction to the microcanonical ensemble average for the same Hamiltonian realization and at the same mean energy $\epsilon _i$,
\begin{equation} \label{eq:sm:eth_equivalence_time}
    \mathcal{Q}^{(\mu)}_{\rm ME} \equiv \frac{1}{\mathcal{N}_{\epsilon _i, \Delta _\epsilon}} \sum _{|E_n^{(\mu)} - \epsilon _i | < \Delta_{\epsilon}} (S_L^z)_{n}, 
\end{equation}
where $\mathcal{N}_{\epsilon _i, \Delta _\epsilon}$ is the normalization factor and the width $\Delta_\epsilon$ of the microcanonical ensemble is sufficiently small. In practice, we either set $\Delta_\epsilon = 5 \times 10^{-2}$, or, if the sum in Eq.~(\ref{eq:sm:eth_equivalence_time}) contains too few elements (which may be the case for small systems), we select $\mathcal{N}_{\epsilon _i, \Delta _\epsilon} = 10$ eigenstates with energies $E_n^{(\mu)}$ closest to $\epsilon _i$, regardless of $\Delta _\epsilon$.
We then define the averaged differences between the diagonal and the microcanonical ensembles via Eq.~(\ref{def_dQ_infty}), which is studied vs $L$ in Fig.~\ref{fig4}(a) of the main text.
The averaging over Hamiltonian realizations in Eq.~(\ref{def_dQ_infty}) is carried out over $N_{\rm samples}=500$ for $L\leq 8$, and the number of realizations is then gradually reduced to $N_{\rm samples} = 500 - 100(L-8)$ for $12 > L\geq 9$. 
For $L=12,13$, we use $N_{\rm samples} = 25, 10$, respectively.

We note that the initial value $\mathcal{Q}(t=0)^{(\mu)}$ at the quantum quench is either $+1/2$ or $-1/2$, depending on the polarization of the spin at $j=L$ in the initial state $|\psi_0\rangle$.
To study properties of time evolution after averaging over different Hamiltonian realizations, we then introduced the spin-spin autocorrelation function $C(t)$ in Eq.~(\ref{def_Ct_autocorr}).
We calculate $C(t)$ by first defining the autocorrelation function for a single Hamiltonian realization,
\begin{equation} \label{def_Ct_mu}
    C(t)^{(\mu)} = \langle \psi_0 | \hat{S}_L^z(t) \hat{S}_L^z (0) | \psi_0\rangle \;,
\end{equation}
which is related to ${\cal Q}(t)^{(\mu)}$ from Eq.~(\ref{def_Qt}) as $C(t)^{(\mu)} = s_L^z {\cal Q}(t)^{(\mu)}$, with $s_L^z = \langle\psi_0|\hat S_L^z |\psi_0\rangle$.
We then obtain $C(t)$ by averaging over different Hamiltonian realizations,
\begin{equation} \label{def_Ct_avr}
    C(t) = {\rm Avr}_\mu\{ C(t)^{(\mu)} \}\;.
\end{equation}
The time evolution of $C(t)$ is shown for $\alpha=0.86$ in Fig.~\ref{fig4}(b) in the main text.

\bibliographystyle{biblev1}
\bibliography{references}
\end{document}